\documentclass[12pt]{article}
\usepackage{amsmath}
\usepackage{graphicx}
\usepackage{rotating}
\usepackage{url, hyperref} 

\newcommand{\blind}{0}

\usepackage{amsmath}
\usepackage{amsfonts,natbib}
\usepackage{amssymb,amsthm}
\usepackage{bm}
\usepackage{mathrsfs,relsize,hyperref}
\usepackage{pdfsync}
\usepackage{graphicx,subfigure}        
\usepackage[top=1.7in, bottom=1.7in, left=0.9in, right=0.9in]{geometry}
\usepackage{color}
\usepackage{lscape}
\usepackage{multirow}
\usepackage{epstopdf}

\allowdisplaybreaks

\begin{document}

\def\spacingset#1{\renewcommand{\baselinestretch}%
{#1}\small\normalsize} \spacingset{1}


\if0\blind
{
  \title{\bf A P\'olya-Gamma Sampler for a Generalized Logistic Regression}
  \author{Luciana Dalla Valle\thanks{ The authors gratefully acknowledge Alex Diana for advising on the PGdraw sampler and Michael W\"{o}gerer for useful feedback on a previous version of the paper. The fourth author is supported by the Chinese Fundamental Research Funds for the Central Universities No 20720181062.}\hspace{.2cm}\\ University of Plymouth, Plymouth, UK\\ Fabrizio Leisen \\ University of Nottingham, Nottingham, UK\\ Luca Rossini\\  Queen Mary University of London, London, UK\\ and\\ Weixuan Zhu\\ Xiamen University, Xiamen, People's Republic of China}
  \maketitle
} \fi

\if1\blind
{
  \bigskip
  \bigskip
  \bigskip
  \begin{center}
    {\LARGE\bf  A P\'olya-Gamma Sampler for a Generalized Logistic Regression}
\end{center}
  \medskip
} \fi

\medskip
\begin{abstract}
In this paper we introduce a novel Bayesian data augmentation approach for estimating the parameters of the generalised logistic regression model.
We propose a P\'olya-Gamma sampler algorithm that allows us to sample from the exact posterior distribution, rather than relying on approximations.
A simulation study illustrates the flexibility and accuracy of the proposed approach to capture heavy and light tails in binary response data of different dimensions.
The methodology is applied to two different real datasets, where we demonstrate that the P\'olya-Gamma sampler provides more precise estimates than the empirical likelihood method, outperforming approximate approaches.
\end{abstract}

\noindent%
\vfill

\newpage
\spacingset{1.5} 
\section{Introduction}
\label{sec:intro}

Data augmentation (DA) is undoubtedly one of the most popular Markov Chain Monte Carlo (MCMC) methods. The idea is to consider the target distribution as the marginal of a joint distribution in an augmented space. Assuming it is easy to sample from the conditional distributions, the algorithm is a straightforward two steps procedure which makes use of a simple Gibbs sampling. Since the seminal work of \cite{Tanner1987}, data augmentation has been extensively studied, developed and used in different contexts. For instance, \cite{Meng1999}, \cite{LiuWu1999}, \cite{Dyk2001} developed strategies to speed up the basic DA algorithm. \cite{Hobert2008} studied the efficiency of DA algorithms. Recently, \cite{Leisen2017} use a DA approach to perform Bayesian inference for the Yule-Simon distribution. The idea is used in the context of infinite hidden Markov models by \cite{HensleyDjuric2017}. The literature on DA is vast and it is difficult to list all the contributions to the field. A review of DA algorithms can be found in \cite{HandbookMCMC}, Chapter 10. 

\medskip

In the context of binary regression, the article of \cite{Albert1993} sets a milestone in the use of DA algorithms. The authors develop exact Bayesian methods for modeling categorical response data by introducing suitable auxiliary variables. In particular, they use a DA approach to estimate the parameters of the probit regression model. However, although some valid approaches have been proposed \citep{Holmes06}, Bayesian inference for the logistic regression was not satisfactorily addressed until the paper of \cite{Polson2013}. The authors proposed an elegant DA approach which makes use of the following identity:
\begin{equation}\label{PG_Identity}
\frac{(e^{\psi})^a}{(1+e^{\psi})^b}=2^{-b}e^{\kappa\psi}\int e^{-\omega\psi^2/2}p(\omega)d\omega
\end{equation}
where $\kappa=a-b/2$, $a > 0$, $b > 0$ and $p(\omega)$ is the density of a PG($b$,0) distribution.  PG($b$,0) is a P\'olya-Gamma distribution with parameters $b$ and 0; we refer to the article of \cite{Polson2013} for details about P\'olya-Gamma distributions. When $\psi=\bm{x^T\beta}$ is a linear function of predictors, the integrand is the kernel of a Gaussian likelihood in $\bm{\beta}$. This naturally suggests a DA scheme which simply requires to be able to sample from the multivariate normal and the P\'olya-Gamma distributions. This algorithm is called the \textit{P\'olya-Gamma sampler}. A theoretical study of the algorithm can be found in \cite{Choi2013}. 

\medskip 

\begin{figure}[h!]
	\centering
	{\includegraphics[width=10cm]{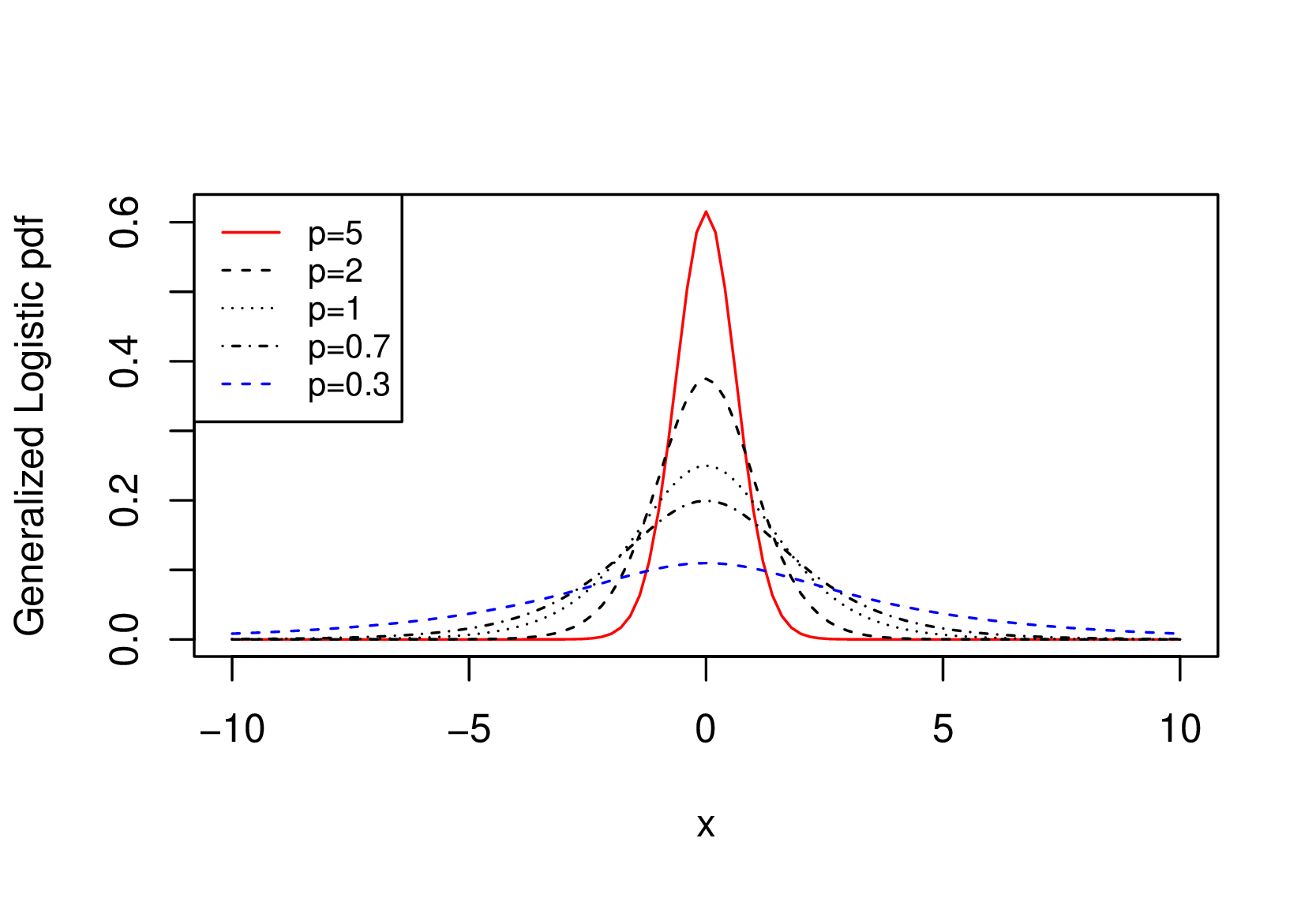}}
	\caption{Probability density function of the generalized logistic with $p=1$, $p=0,3$, $p=0,7$, $p=2$ and $p=5$.}
	\label{GenLog_pdf}
\end{figure}

In this paper, we propose a P\'olya-Gamma sampler to estimate the parameters of the generalized logistic regression model introduced in \cite{OurArticle2019}. The authors studied immigration in Europe employing a novel regression model which makes use of a generalized logistic distribution. Their work is motivated by the need of a logistic model with heavy tails. The distribution considered in the paper has the following density function:
\begin{equation}\label{GLogist}
f(x)=\frac{1}{B(p,p)}\frac{e^{px}}{\left(1+e^x\right)^{2p}} \quad x\in \mathbb{R},
\end{equation}
where the parameter $p$ controls the tails of the distribution. The standard logistic density can be recovered with $p=1$.  Compared with the standard logistic distribution, $p>1$ means lighter tails and $0<p<1$ means heavier tails, as shown in Figure \ref{GenLog_pdf}. Unfortunately, the distribution function is not explicit:
\begin{equation}\label{GLogistDF}
F(x)=\frac{1}{B(p,p)} B\left(\frac{e^x}{1+e^x};p,p\right) \quad x\in \mathbb{R} 
\end{equation}
where $B(t;p,q)=\int_0^t x^{p-1}(1-x)^{q-1}dx$, with $0<t<1$, is the incomplete Beta function. Therefore, a straightforward use of the methodology of \cite{Polson2013} is not possible. To overcome the problem, \cite{OurArticle2019} use an approximate Bayesian computational method that relies on the empirical likelihood (see \cite{Mengersen13} and \cite{karabatsos2018}). However, since the empirical likelihood approach is based on an approximation of the posterior distribution, this method may lead to wide credible intervals and sometimes ambiguous estimates, especially for the tail parameter. 

\medskip

In this paper, we propose a novel approach for estimating the generalised logistic regression, that allows us to draw samples from the exact posterior distribution and is able to overcome the drawbacks of the empirical likelihood method. 
We show how to set a DA scheme for the generalized logistic regression which makes use of the P\'olya-Gamma identity in equation \eqref{PG_Identity}. 
We test the performance of the proposed approach on simulated data and on two different real dataset: the first dataset includes people's opinions towards immigration in Europe; the second contains information about  the recidivism of criminals detained in Iowa.
We show that the P\'olya-Gamma sampler yields accurate results in high-dimension and outperforms the empirical likelihood algorithm providing a more precise estimation of the model parameters.

The rest of the paper is organised as follows. In Section 2, we introduce the generalized logistic regression framework in a Bayesian setting. In Section 3, we provide the details of the P\'olya-Gamma sampler for this model.
Section 4 and Section 5 illustrate the algorithm performance with simulated and real data. Section 6 concludes. 

\section{The Generalized Bayesian Logistic Regression} \label{GBLR}

%
%
%
%

Consider a binary regression set-up in which $Y_1,\dots,Y_n$ are independent Bernoulli random variables such that  $\Pr(Y_i=1|\bm{\beta})=H(x_i^T\bm{\beta})$ where $x_i$ is a $k\times 1$ vector of known covariates associated with $Y_i$, $i = 1, \ldots, n$, $\bm{\beta}$ is a $k\times 1$ vector of unknown regression coefficients, and $H:\mathbb{R}\rightarrow (0,1)$ is a distribution function.
In this case, the likelihood function is given by 
\begin{equation}\label{Likelihood}
L(\bm{\beta})=\prod_{i=1}^n [H(x_i^T\bm{\beta})]^{y_i}[1-H(x_i^T\bm{\beta})]^{1-y_i}
\end{equation}
If $H(x)=\Phi(x)$ is the distribution function of a Gaussian, then  we are in the probit regression framework.  If $H(x)=e^x(1+e^x)^{-1}$, then we are in the special case of the standard logistic regression.  \cite{Albert1993} proposed a DA approach for sampling from the posterior distribution of the probit regression model. Their method makes use of auxiliary variables which allow an easy implementation of the Gibbs sampler. \cite{Polson2013} proposed an elegant algorithm for tackling the logistic regression case.  

The generalized logistic distribution in equation \eqref{GLogist} has mean zero and scale one. Consider now a generalized logistic with mean $x_i^T\bm{\beta}$ and scale one, i.e. with probability density function:
\begin{equation}\label{GLogistNonzero}
f(x)=\frac{1}{B(p,p)}\frac{e^{p(x-x_i^T\bm{\beta})}}{\left(1+e^{(x-x_i^T\bm{\beta})}\right)^{2p}} \quad x\in \mathbb{R} 
\end{equation}
We denote the above distribution with $\mbox{GLog}(x_i^T\bm{\beta},1)$. Mimicking \cite{Albert1993}, suppose to sample $Z_i$ from a $\mbox{GLog}(x_i^T\bm{\beta},1)$, $i=1,\dots,n$, and set $Y_i=1$ if $Z_i>0$. Otherwise, if  $Z_i\leq 0$, then set $Y_i=0$.  It is easy to see that 
$$P(Y_i=1)=\frac{1}{B(p,p)} B\left(\frac{e^{x_i^T\bm{\beta}}}{1+e^{x_i^T\bm{\beta}}};p,p\right).$$
The joint posterior distribution is
\begin{align}\label{Post}
\pi(\bm{\beta},p,\bm{Z}|\bm{y})&\propto \pi(\bm{\beta})\pi(p)\prod_{i=1}^n \left\{\mathbb{I}\left(Z_i>0\right)\mathbb{I}(y_i=1)+\mathbb{I}\left(Z_i\le 0\right)\mathbb{I}(y_i=0)\right\}\nonumber\\
&\hspace{6cm}\times \frac{1}{B(p,p)}\frac{e^{p(Z_i-x_i^T\bm{\beta})}}{\left(1+e^{(Z_i-x_i^T\bm{\beta})}\right)^{2p}}
\end{align}
where $\pi(\bm{\beta})$ and $\pi(p)$ are the prior distributions of $\beta$ and $p$, respectively. With the probit model, the technique of \cite{Albert1993} works because the full conditional distributions can be resorted to normal distributions (or truncated normal distributions) which are easy to sample from. In our case, the usual normal prior would lead to a full conditional on $\beta$ which is not explicit, requiring the use of a Metropolis step (or alternative algorithms).  The P\'olya-Gamma identity in \eqref{PG_Identity} allows us to overcome this problem. In particular, we get
$$\frac{e^{p(Z_i-x_i^T\bm{\beta})}}{\left(1+e^{(Z_i-x_i^T\bm{\beta})}\right)^{2p}}=2^{-2p}\int_0^{+\infty}e^{-\omega_i(Z_i-x_i^T\bm{\beta})^2/2}p(\omega_i)d\omega_i,$$
where $p(\omega)$ is a P\'olya-Gamma, $\mbox{PG}(2p,0)$, distribution. Note that $e^{-\omega(x_i^T\bm{\beta})^2/2}p(\omega)$ is the unnormalized density of a $\mbox{PG}(2p,x_i^T\bm{\beta})$ random variable. Therefore, the augmented posterior distribution is 
\begin{align}\label{AugPost}
\pi(\bm{\beta},p,\bm{Z},\bm{\omega}|\bm{y})&\propto \pi(\bm{\beta})\pi(p)\prod_{i=1}^n \left\{\mathbb{I}\left(Z_i>0\right)\mathbb{I}(y_i=1)+\mathbb{I}\left(Z_i\le 0\right)\mathbb{I}(y_i=0)\right\}\nonumber\\
&\hspace{6cm}\times 2^{-2p} e^{-\omega_i(Z_i-x_i^T\bm{\beta})^2/2}p(\omega_i).
\end{align}

\medskip

\section{Inference Sampling Strategy} \label{Inference}

The algorithm proposed in \cite{OurArticle2019} makes use of an approximate likelihood, which relies on an approximation of the posterior distribution (see \cite{Mengersen13} and \cite{ZhuLeisen2016}). The algorithm presented in this paper allows us to sample from the true posterior distribution, making it more appealing to perform the Bayesian inferential exercise. 

Starting from the augmented posterior distribution in \eqref{AugPost}, this Section introduces the full conditional distributions required to run the algorithm. The variables involved are
$$\bm{Z},\bm{\omega}, \bm{\beta}, p,$$
where $\bm{Z}=(Z_1,\dots,Z_n)$, $\bm{\omega}=(\omega_1,\dots,\omega_n)$ and $\bm{\beta}=(\beta_1,\dots,\beta_k)$.

\medskip

We will focus on two cases: 1) the parameter $p$ is known, 2) the parameter $p$ is unknown. In the first case, 
the algorithm performance is very good and accurate. In the second case, the estimation of the parameter $p$ introduces more noise in the algorithm leading to a less accurate, but still acceptable, estimation. Section 4 will provide a simulation study for both scenarios. The real data analysis assumes that $p$ is unknown. 

\subsection*{The parameter $p$ is known}

In this case, the posterior distribution in \eqref{AugPost} reduces to: 
\begin{align}\label{AugPostWP}
\pi(\bm{\beta},\bm{Z},\bm{\omega}|\bm{y})&\propto \pi(\bm{\beta})\prod_{i=1}^n \left\{\mathbb{I}\left(Z_i>0\right)\mathbb{I}(y_i=1)+\mathbb{I}\left(Z_i\le 0\right)\mathbb{I}(y_i=0)\right\}\nonumber\\
&\hspace{6cm}\times 2^{-2p} e^{-\omega_i(Z_i-x_i^T\bm{\beta})^2/2}p(\omega_i).
\end{align}
The full conditional distributions are derived below and the variables of interest are: 
$$\bm{Z},\bm{\omega}, \bm{\beta}.$$

\noindent\textbf{Full conditional for $Z_i$.} The random variables $Z_1,\dots,Z_n$ are independent with full conditional distributions:
\begin{align}
\pi(Z_i|\bm{y},\bm{\beta},\bm{\omega})&\propto \left\{\mathbb{I}\left(Z_i>0\right)\mathbb{I}(y_i=1)+\mathbb{I}\left(Z_i\le 0\right)\mathbb{I}(y_i=0)\right\} e^{-\omega_i(Z_i-x_i^T\bm{\beta})^2/2} \notag \\
&\propto \left\{\mathbb{I}\left(Z_i>0\right)\mathbb{I}(y_i=1)+\mathbb{I}\left(Z_i\le 0\right)\mathbb{I}(y_i=0)\right\} e^{-\frac{\omega_i}{2}(Z_i^2-2Z_i x_i^T\bm{\beta})} \notag
\end{align}
which are truncated normal distributions. More precisely,
\begin{itemize}
	\item if $y_i=1$, then  $\pi(Z_i|\bm{y},\bm{\beta},\bm{\omega})$ is distributed as a $\mathcal{N}(x_i^T\bm{\beta},\omega_i^{-1}) $ truncated to the left by $0$;
	\item if $y_i=0$, then $\pi(Z_i|\bm{y},\bm{\beta},\bm{\omega})$ is distributed as a  $\mathcal{N}(x_i^T\bm{\beta},\omega_i^{-1}) $ truncated to the right by $0$.
\end{itemize}

\medskip

\noindent\textbf{Full conditional for $\omega_i$.} Following \cite{Polson2013}, it is easy to see that the full conditional distribution of $\omega_i$ is:
\begin{align*}
\pi(\omega_i|\bm{\beta},\bm{Z},\bm{y}) \propto e^{-\omega_i(Z_i-x_i^T\bm{\beta})^2/2}p(\omega_i) \sim \mbox{PG}\left(2p, Z_i - x_i^{T}\bm{\beta}\right) 
\end{align*}
An efficient method for sampling P\'olya-Gamma random variables is implemented in a modified version of the R package \texttt{BayesLogit} \citep{Polson_RCode_2013}.

\medskip

\noindent \textbf{Full conditional for $\bm{\beta}$.}
Let $\Omega=\mbox{diag}(\omega_1,\dots,\omega_n)$. Assuming a multivariate $\mathcal{N}(\bm{\beta}^*,B^*)$ prior for $\bm{\beta}$, the full conditional distribution of $\bm{\beta}$ is
\begin{align*}
\pi(\bm{\beta}|\mathbf{Z},\bm{y},\bm{\omega}) &\propto \pi(\bm{\beta})\prod_{i=1}^n \exp{\left\{-\frac{\omega_i}{2}(Z_i-x_i^T\bm{\beta})^2\right\}} \\
&= \exp{\left\{-\frac{1}{2}\left[ \bm{\beta}^T\left(X^T \Omega X +B^{*-1}\right) \bm{\beta}-2\bm{\beta}^T \left(X^T \Omega Z + B^{*-1}\bm{\beta}^*\right)\right]\right\}} \\
&\sim \mathcal{N}(\tilde{\bm{\beta}}^*,\tilde{V}_{\beta}^*)
\end{align*}
where 
\begin{align*}
\tilde{\bm{\beta}}^*&=\left(X^T\Omega X + B^{*-1}\right)^{-1} \left(X^T \Omega Z + B^{*-1}\bm{\beta}^*\right) \\
\tilde{V}_{\beta}^*&=\left(X^T \Omega X + B^{*-1}\right)^{-1}. 
\end{align*}

\subsection*{The parameter $p$ is unknown}

\medskip

In this case, the full conditional for $p$ is:
\begin{align*}
\pi(p|\bm{\beta},\bm{Z},\bm{\omega},\bm{y})&\propto \pi(p)\prod_{i=1}^n \frac{2^{-2p}}{B(p,p)}p(\omega_i)
\end{align*}
where $p(\omega_i)$ is a $\mbox{PG}(2p,0)$ and we assume a Gamma prior for $p$. Sampling from this full conditional distribution is not straightforward and alternative strategies must be employed. One can marginalize the above full conditional by integrating out the $\omega$'s and applying the slice sampling of \cite{neal2003}. Alternatively, one can further integrate out the $\bm{Z}$ and obtain the following conditional
\begin{align}
	\pi(p|\bm{\beta},\bm{y})&\propto \pi(p)\prod_{i=1}^n
	\left[ \frac{1}{B(p,p)} B\left(\frac{e^{x_i^T\bm{\beta}}}{1+e^{x_i^T\bm{\beta}}};p,p\right)\right]^{y_i} \left[1- \frac{1}{B(p,p)} B\left(\frac{e^{x_i^T\bm{\beta}}}{1+e^{x_i^T\bm{\beta}}};p,p\right)\right]^{1-y_i}
\end{align}
 The slice sampling algorithm requires the computation of the inverse of the above full conditional distribution. We used the function \texttt{uniroot} available in R to compute the numerical inverse. 

The next sections illustrate the performance of the algorithm with simulated and real data. 
\section{Simulation Studies} \label{Simu}
In this section, we assess the performance of the algorithm by implementing different simulation experiments. In particular, we use four different values of the tail parameter $p$: $p=0.3$ and $p=0.7$ (heavy tails) and $p=1.5$ and $p=3$ (light tails). We focus on two scenarios: 

\begin{enumerate}
	\item $\bm{\beta} = (1,-1,-3,1,3)$, with $k = 5$;
	\item $\bm{\beta} = (2.3,1,-2,1.5,-2.7,0.2,-1.4,3,-0.6,-1.2)$, with $k = 10$.
\end{enumerate}

\noindent We assume a standard normal distribution for the explanatory variables. For each scenario and each value of $p$, we generate $100$ different datasets of sample sizes $n=100$ and $n=250$, respectively. We analyse two different cases: the case with $p$ known and the case with $p$ unknown. 

\subsection{Case with $p$ known}

We considered a vague prior for $\bm{\beta}$, such as the multivariate normal distribution, $\mathcal{N}(\bm{\nu},B)$, with prior mean vector $\bm{\nu} = 0$ and prior covariance matrix $B = 100 \cdot I_k$.
For each of the $100$ different simulated datasets, we run $20,000$ iterations of the MCMC algorithm and discarded the first $5,000$ iterations as burn-in period. 
Table \ref{beta5_dim_fix} lists the posterior means of the five-dimensional $\bm{\beta}$ of scenario 1 for the different values of $p$ and $n$, averaged over the $100$ simulated datasets. The values in brackets show the standard deviations over the $100$ simulations. We notice that the posterior means of $\bm{\beta}$ are very close to the true values, with a better estimation performance as $n$ increases from $100$ to $250$. 
Table \ref{beta10_dim_fix} shows the posterior means and, in brackets, the standard deviations over the $100$ simulations for the ten-dimensional scenario. These results are in line with the five-dimensional case, thus leading us to conclude that increasing the number of covariates is not an issue in terms of estimation accuracy. Moreover, the standard deviations in Tables \ref{beta5_dim_fix} and \ref{beta10_dim_fix} substanstially decrease for both scenarios when we move from $n=100$ to $n=250$. 

\noindent We compare the performance of our P\'olya-Gamma sampler to the one of the approximate Bayesian computation with empirical likelihood ($BC_{el}$) adopted in \cite{OurArticle2019}. Table \ref{EL:beta5_dim_fix} and Table \ref{EL:beta10_dim_fix} display the posterior means and the standard deviations over 100 simulation datasets estimated with $BC_{el}$. The sampling efficiency of $BC_{el}$ depends heavily on the prior distribution. So we consider a normal prior distribution $\mathcal{N}(\bm{\nu}=0,B=9\cdot I_k)$ for $\bm{\beta}$. The results are based on 20,000 posterior samples obtained with $BC_{el}$. From the table, one can see that both methods achieve similar accuracy level in terms of the estimation of the posterior means. However, the standard deviations given by $BC_{el}$ is larger than the ones obtained by P\'olya-Gamma sampler, especially when $p$ parameter is large. Finally, consider that the prior distribution used by P\'olya-Gamma sampler is much vaguer than the one used by $BC_{el}$, we conclude that P\'olya-Gamma sampler outperforms $BC_{el}$.

\begin{figure}[h!]
	\centering
	\begin{tabular}{ccc}
		{\includegraphics[width=4.75cm]{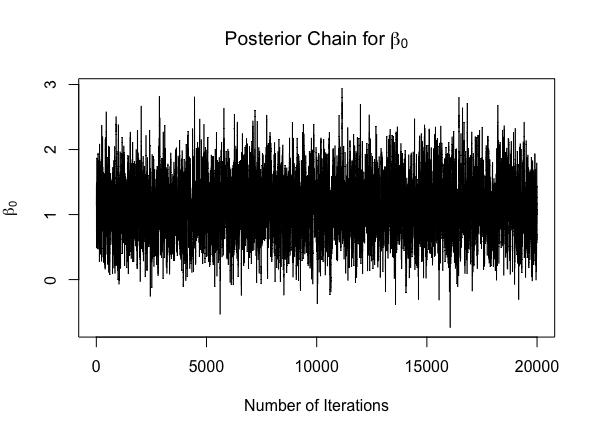}} &
		{\includegraphics[width=4.75cm]{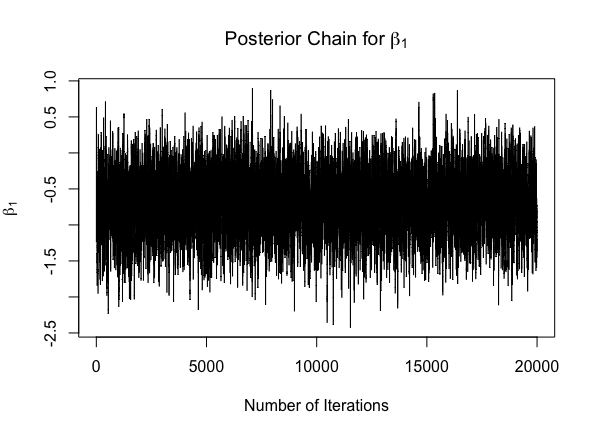}} &
		{\includegraphics[width=4.75cm]{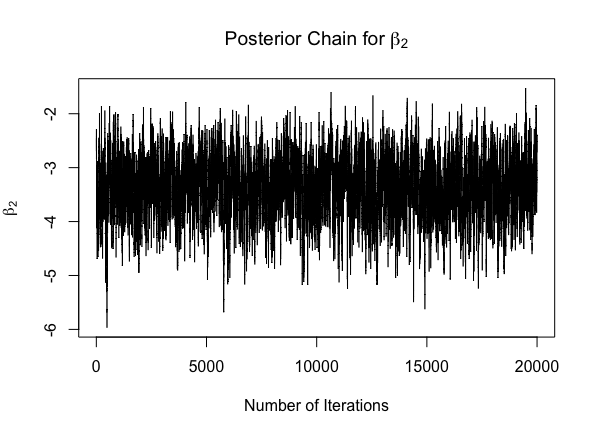}} \\
		{\includegraphics[width=4.75cm]{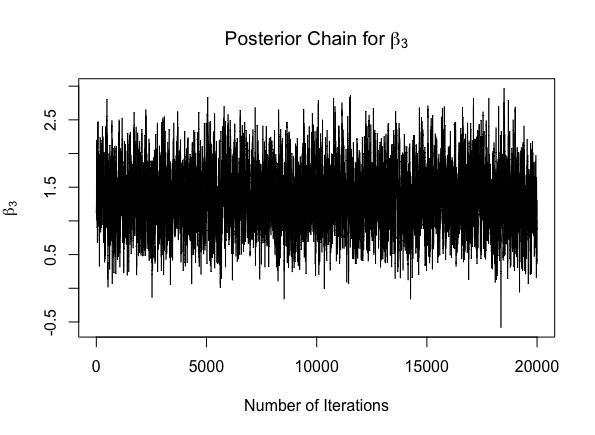}} &
		{\includegraphics[width=4.75cm]{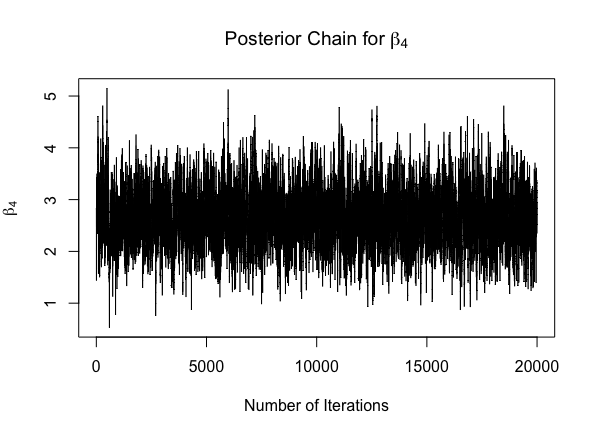}} \\
	\end{tabular}
	\caption{Case with $p$ known: sample chains of the parameters posterior distributions for the simulated data with sample size $n=250$ and with true parameter values $p=0.3$ and $\bm{\beta} = (\beta_0,\beta_1,\beta_2,\beta_3,\beta_4) = (1,-1,-3,1,3)$.}
	\label{Fig_Chain_Simu_fix}
\end{figure}

\noindent In order to illustrate the complete inferential procedure, we show how the model parameters are estimated. Figure \ref{Fig_Chain_Simu_fix} displays the posterior chains for a specific dataset with sample size $n=250$, where $p=0.3$ and $\bm{\beta} = (\beta_0,\beta_1,\beta_2,\beta_3,\beta_4) = (1,-1,-3,1,3)$. As shown in Figure \ref{Fig_Chain_Simu_fix}, the chains, after a burn-in of $5,000$ iterations, converge quickly to the true values of all parameters and give an excellent representation of the real parameter values.

\noindent We also performed a convergence analysis for scenario 1, with five-dimensional $\bm{\beta}${\footnote{We performed other convergence tests for the ten-dimensional case. These results are omitted for lack of space but are available on request.}}. The analysis was carried out using the R \texttt{coda} package \citep{Plummer2006}. In particular, we used single datasets of sample sizes $n=100$ and $n=250$, respectively, where we fixed $p=0.3$. We computed Geweke's convergence tests and the autocorrelation and partial autocorrelation functions for the values of $\bm{\beta}$. Table \ref{Tab_Gew_fix} reports the results of Geweke's convergence test \citep{Geweke1992}, indicating no issues since the absolute values of the test statistics are all lower than $1.96$.

\begin{table}[htbp]
	\centering
	\begin{tabular}{lcccccc}
		\hline
		& $p$ &  $\beta_0$ & $\beta_1$ & $\beta_2$ &  $\beta_3$ &  $\beta_4$  \\
		\hline
		\emph{true values} & 0.3 & 1 & -1 & -3 & 1 & 3 \\
		\hline
		$n = 100$   & -- & 1.113 &  -1.128 &  -3.191 &  1.056 &  3.396\\
& -- &  (0.795) & (0.666) & (0.937) & (0.696) & (0.904) \\		
		$n = 250$  &  -- & 1.041 & -1.061 & -3.173 & 1.027 & 3.273\\
		& -- & (0.411) & (0.412) & (0.548) & (0.451) & (0.555) \\
		\hline
		\emph{true values} & 0.7 & 1 & -1 & -3 & 1 & 3 \\
		\hline
				$n = 100$   & -- & 1.179 & -1.094 &  -3.282 &  1.148  & 3.309\\
& -- & (0.448) & (0.545) & (0.640) & (0.558) & (0.659) \\
		$n = 250$   & -- & 1.029 & -1.110 &  -3.187 &  1.085 &  3.206 \\
& -- & (0.304) & (0.295) & (0.503) & (0.312) & (0.479) \\
		\hline
		\emph{true values} & 1.5 & 1 & -1 & -3 & 1 & 3 \\
		\hline
		$n = 100$   & -- & 1.169  & -1.124  & -3.605  & 1.158 &  3.582\\
& -- & (0.426) & (0.430) & (0.867) & (0.470) & (0.782) \\
		$n = 250$    & -- &  1.096 &  -1.068 &  -3.277 &  1.111 &  3.326\\
& -- & (0.258) & (0.238) & (0.541) & (0.298) & (0.518) \\
		\hline
		\emph{true values} & 3 & 1 & -1 & -3 & 1 & 3 \\ 
		\hline
			$n = 100$   &  -- & 1.183 &  -1.214 &  -3.681 & 1.239 &  3.729 \\
& -- &  (0.417) & (0.391) & (0.853) & (0.478) & (0.984) \\
		$n = 250$   & -- & 1.130  & -1.154 &  -3.431  & 1.119 &  3.408\\
& -- & (0.253) & (0.284) & (0.653) & (0.239) & (0.662) \\
		\hline
	\end{tabular}
	\caption{PG sampler. Case with $p$ known: posterior means over the $100$ different simulated datasets for $n$ equal to $100$ and $250$ with true values of $p = 0.3$ (first panel), $p=0.7$ (second panel), $p=1.5$ (third panel) and $p=3$ (fourth panel) and $(\beta_0,\beta_1,\beta_2,\beta_3,\beta_4) = (1,-1,-3,1,3)$. The values in brackets are the standard deviations over the $100$ different simulations.}
	\label{beta5_dim_fix}
\end{table}

\begin{table}[htbp]
	\centering
	\begin{tabular}{lcccccc}
		\hline
		& $p$ &  $\beta_0$ & $\beta_1$ & $\beta_2$ &  $\beta_3$ &  $\beta_4$  \\
		\hline
		\emph{true values} & 0.3 & 1 & -1 & -3 & 1 & 3 \\
		\hline
		$n = 100$   & -- & 1.185 &	-1.136& 	-3.356& 	1.159& 	3.403 \\
		& -- &  (0.773)& 	(0.708) &	(0.900) &	(0.785) &	(0.844 ) \\		
		$n = 250$  &  -- & 1.052 &	-1.067& 	-3.185& 	1.045& 	3.292 \\
		& -- & (0.506) & (0.435) & (0.600) & (0.488) & (0.630) \\
		\hline
		\emph{true values} & 0.7 & 1 & -1 & -3 & 1 & 3 \\
		\hline
		$n = 100$   & -- & 1.146 & -1.177 &  -3.751 &  1.331  & 3.65\\
		& -- & (0.859) & (0.711) & (0.750) & (0.778) & (0.819) \\
		$n = 250$   & -- & 1.315 & -1.214 &  -3.289 &  1.251 &  3.304 \\
		& -- & (0.511) & (0.412) & (0.614) & (0.407) & (0.687) \\
		\hline
		\emph{true values} & 1.5 & 1 & -1 & -3 & 1 & 3 \\
		\hline
		$n = 100$   & -- & 1.326& 	-1.086& 	-3.747& 	1.187& 	3.714 \\
		& -- & (1.214) & (0.618) & (1.382) & (0.600) & (0.997) \\
		$n = 250$    & -- & 1.064& 	-1.019& 	-3.068& 	1.010& 	3.643  \\
		& -- & (0.513) & (0.376) & (0.598) & (0.399) & (0.736) \\
		\hline
		\emph{true values} & 3 & 1 & -1 & -3 & 1 & 3 \\ 
		\hline
		$n = 100$   &  -- &  1.473 &	-1.373& 	-3.761& 	1.373& 	3.982 \\
		& -- &  (1.044) & (1.049) & (1.556) & (0.913) & (1.739) \\
		$n = 250$   & -- & 1.180 &	-1.256& 	-3.601& 	1.186& 	3.534 \\
		& -- & (0.513) & (0.591) & (1.210) & (0.539) & (1.111) \\
		\hline
	\end{tabular}
	\caption{$BC_{el}$ sampler. Case with $p$ known: posterior means over the $100$ different simulated datasets for $n$ equal to $100$ and $250$ with true values of $p = 0.3$ (first panel), $p=0.7$ (second panel), $p=1.5$ (third panel) and $p=3$ (fourth panel) and $(\beta_0,\beta_1,\beta_2,\beta_3,\beta_4) = (1,-1,-3,1,3)$. The values in brackets are the standard deviations over the $100$ different simulations.}
	\label{EL:beta5_dim_fix}
\end{table}

\begin{sidewaystable}[htbp]
		\centering
		\begin{tabular}{lccccccccccc}
			\hline
			& $p$ &  $\beta_0$ & $\beta_1$ & $\beta_2$ &  $\beta_3$ &  $\beta_4$ &  $\beta_5$ &  $\beta_6$ &  $\beta_7$ &  $\beta_8$ &  $\beta_9$   \\
			\hline
			\emph{true values} & 0.3 & 2.3 & 1 & -2 & 1.5 & -2.7 & 0.2 & -1.4 & 3 & -0.6 & -1.2\\
			\hline
			$n = 100$   & -- & 2.597 &  1.049  & -2.401  & 1.737  & -2.945 &  0.063 &  -1.617 &  3.366  & -0.785 &  -1.369 \\
  & -- &  (0.760) & (0.894) & (0.993) & (0.766) & (0.836) & (1.001) & (0.806) & (0.960) & (0.762) & (0.800) \\
			$n = 250$  & -- & 2.479  & 1.098  & -2.243  & 1.733  & -3.035 &  0.312  & -1.563  & 3.374 &  -0.709 &  -1.259 \\
& -- & (0.569) &  (0.445) &  (0.481) &  (0.539) &  (0.650) &  (0.507) &  (0.527) &  (0.627) &  (0.542) &  (0.486) \\
			\hline
			\emph{true values} & 0.7 & 2.3 & 1 & -2 & 1.5 & -2.7 & 0.2 & -1.4 & 3 & -0.6 & -1.2\\
			\hline
		$n = 100$ & -- & 2.925 &  1.361 &   -2.386 &   1.914 &   -3.466 &   0.287 &   -1.736  &  3.788  &  -0.752  &  -1.539\\
  & -- & (0.719) & (0.658) & (0.754) & (0.619) & (0.861) & (0.570) & (0.697) & (0.832) & (0.606) & (0.680) \\			
  $n = 250$  &  -- & 2.573  &1.098  &-2.230  &1.686  &-3.072 & 0.179 & -1.574  &3.369  &-0.667  &-1.348 \\
  & -- & (0.475) & (0.370) & (0.398) & (0.459) & (0.528) & (0.310) & (0.385) & (0.505) & (0.313) & (0.384) \\
			\hline
			\emph{true values} & 1.5 & 2.3 & 1 & -2 & 1.5 & -2.7 & 0.2 & -1.4 & 3 & -0.6 & -1.2\\
			\hline
			$n = 100$  & -- &  3.126 &  1.401  & -2.903  & 2.049  & -3.784 &  0.311 &  -1.977 &  4.220  & -0.874 &  -1.679 \\
   & -- & (0.678) & (0.583) & (0.691) & (0.657) & (0.767) & (0.528) & (0.671) & (0.918) & (0.554) & (0.552) \\
   		$n = 250$  & -- & 2.816 &  1.180  & -2.440  & 1.810  & -3.265  & 0.248  & -1.735 & 3.623  & -0.680  & -1.428 \\
   & -- & (0.549) & (0.321) & (0.517) & (0.390) & (0.635) & (0.264) & (0.376) & (0.620) & (0.276) & (0.378) \\
			\hline
			\emph{true values} & 3 & 2.3 & 1 & -2 & 1.5 & -2.7 & 0.2 & -1.4 & 3 & -0.6 & -1.2\\ 
			\hline
			$n = 100$  & -- & 3.442 &  1.600  & -3.048  & 2.171  & -4.101  & 0.227  & -2.225  & 4.558  & -1.021 &  -1.779\\
& -- &  (0.672) & (0.607) & (0.739) & (0.577) & (0.764) & (0.554) & (0.673) & (0.870) & (0.517) & (0.517) \\		
	$n = 250$  &  -- &  3.005   & 1.236   & -2.594  &  1.954   & -3.468   & 0.291   & -1.833   & 3.864   & -0.795   & -1.536 \\
& -- & (0.609) & (0.297) & (0.546) & (0.455) & (0.750) & (0.263) & (0.483) & (0.734) & (0.309) & (0.359) \\
			\hline
		\end{tabular}
		\caption{PG sampler. Case with $p$ known: posterior means over the $100$ different simulated datasets for $n$ equal to $100$ and $250$ with true values of $p = 0.3$ (first panel), $p=0.7$ (second panel), $p=1.5$ (third panel) and $p=3$ (fourth panel) and $(\beta_0,\beta_1,\beta_2,\beta_3,\beta_4,\beta_5,\beta_6,\beta_7,\beta_8,\beta_9) = (2.3,1,-2,1.5,-2.7,0.2,-1.4,3,-0.6,-1.2)$. The values in brackets are the standard deviations over the $100$ different simulations.}
		\label{beta10_dim_fix}
\end{sidewaystable}

\begin{sidewaystable}[htbp]
	\centering
	\begin{tabular}{lccccccccccc}
		\hline
		& $p$ &  $\beta_0$ & $\beta_1$ & $\beta_2$ &  $\beta_3$ &  $\beta_4$ &  $\beta_5$ &  $\beta_6$ &  $\beta_7$ &  $\beta_8$ &  $\beta_9$   \\
		\hline
		\emph{true values} & 0.3 & 2.3 & 1 & -2 & 1.5 & -2.7 & 0.2 & -1.4 & 3 & -0.6 & -1.2\\
		\hline
		$n = 100$   & -- &2.868 &	1.058& 	-2.617& 	1.379& 	-3.035& 	-0.178& 	-1.687& 	3.656& 	-0.920 &	-0.980\\
		& -- &  (1.046) & (1.282) & (1.527) & (1.381) & (1.500) & (1.488) & (1.493) & (1.619) & (1.261) & (1.463) \\
		$n = 250$  & -- & 2.875 &	1.221 &	-2.495 &	1.843& 	-3.235 &	0.068 &	-1.880& 	3.693 &	-0.772& 	-1.116  \\
		& -- & (1.239) &  (1.206) &  (1.137) &  (1.333) &  (1.579) &  (1.199) &  (1.343) &  (1.231) &  (1.145) &  (1.313) \\
		\hline
		\emph{true values} & 0.7 & 2.3 & 1 & -2 & 1.5 & -2.7 & 0.2 & -1.4 & 3 & -0.6 & -1.2\\
		\hline
		$n = 100$ & -- & 3.145 &  1.089 &   -2.975 &   1.874 &   -3.429 &   0.168 &   -1.879 &  3.598  &  -0.818  &  -1.497\\
		& -- & (1.819) & (1.674) & (1.816) & (1.898) & (1.901) & (1.753) & (1.852) & (2.011) & (1.844) & (1.910) \\			
		$n = 250$  &  -- & 2.855  &1.150  &-2.331  &1.716  &-3.227 & 0.149 & -1.674  &3.445  &-0.750  &-1.421 \\
		& -- & (1.508) & (1.501) & (1.389) & (1.585) & (1.783) & (1.501) & (1.441) & (1.612) & (1.431) & (1.451) \\
		\hline
		\emph{true values} & 1.5 & 2.3 & 1 & -2 & 1.5 & -2.7 & 0.2 & -1.4 & 3 & -0.6 & -1.2\\
		\hline
		$n = 100$  & -- & 4.270 &	0.988 &	-1.849 &	1.139& 	-1.806 &	-0.181 &	-0.929& 	2.260& 	-0.662& 	-0.826 \\
		& -- & (4.033) & (2.160) & (2.320) & (2.242) & (2.781) & (2.282) & (2.552) & (2.663) & (2.032) & (1.611) \\
		$n = 250$  & -- & 3.111& 	1.244 &	-2.424& 	1.595& 	-3.861& 	0.409 &	-1.883 &	4.051 &	-0.184 &	-1.593  \\
		& -- & (1.083) & (1.159) & (1.272) & (1.441) & (1.571) & (1.407) & (1.422) & (1.422) & (1.264) & (1.186) \\
		\hline
		\emph{true values} & 3 & 2.3 & 1 & -2 & 1.5 & -2.7 & 0.2 & -1.4 & 3 & -0.6 & -1.2\\ 
		\hline
		$n = 100$  & -- & 2.823 &	0.301 &	-0.607& 	0.790& 	-0.966 &	0.211 &	-0.270 &	0.907& 	0.015& 	-0.125 \\
		& -- &  (3.488) & (2.364) & (2.492) & (2.319) & (2.964) & (2.425) & (2.510) & (2.449) & (2.365) & (2.791) \\		
		$n = 250$  &  -- &  2.933 &	1.079 &	-1.769 &	1.670 &	-2.906 &	0.271 &	-1.350 &	3.035 &	-0.830 &	-1.172 \\
		& -- & (1.576) & (1.431) & (1.929) & (1.660) & (1.684) & (1.687) & (1.740) & (1.968) & (1.352) & (1.888) \\
		\hline
	\end{tabular}
	\caption{$BC_{el}$ sampler.Case with $p$ known: posterior means over the $100$ different simulated datasets for $n$ equal to $100$ and $250$ with true values of $p = 0.3$ (first panel), $p=0.7$ (second panel), $p=1.5$ (third panel) and $p=3$ (fourth panel) and $(\beta_0,\beta_1,\beta_2,\beta_3,\beta_4,\beta_5,\beta_6,\beta_7,\beta_8,\beta_9) = (2.3,1,-2,1.5,-2.7,0.2,-1.4,3,-0.6,-1.2)$. The values in brackets are the standard deviations over the $100$ different simulations.}
	\label{EL:beta10_dim_fix}
\end{sidewaystable}

\begin{table}[htbp]
	\centering
	\begin{tabular}{llcc|llc}
		\hline
		& Variables & Test Statistic & &  & Variables & Test Statistic\\
		\hline
		$n = 100$ & $\beta_0 = 1$ & -0.8422 & & $n = 250$ & $\beta_0 = 1$ &-0.0932  \\
		& $\beta_1 = -1$ & -0.6155 & & & $\beta_1 = -1$ & 0.9662 \\
		& $\beta_2 = -3$ &  0.2379  & & & $\beta_2 = -3$ & 0.4544\\
		& $\beta_3 = 1$ & -0.2488  & & & $\beta_3 = 1$ &   0.4355 \\
		& $\beta_4 = 3$ & 0.4159 & & & $\beta_4 = 3$ &    -0.3423 \\
		\hline
	\end{tabular}
	\caption{Case with $p$ known: Geweke's test statistics for the posterior chain of the unknown coefficients, $\bm{\beta}= (\beta_0,\beta_1,\beta_2,\beta_3,\beta_4)$ for a simulated dataset of sample size $n=100$ (left panel) and $n=250$ (right panel) with $p=0.3$.}
	\label{Tab_Gew_fix}
\end{table}

\subsection{Case with $p$ unknown}

We considered vague priors for $\bm{\beta}$ and $p$. In particular, a multivariate normal distribution, $\mathcal{N}(\bm{\nu},B)$, is chosen for $\bm{\beta}$ with prior mean vector $\bm{\nu} = 0$ and prior covariance matrix $B = 100 \cdot I_k$. For $p$ we chose a gamma prior, $\mathcal{G}a(a,b)$ with hyperparameters $a=b=1$. For each of the $100$ different simulated datasets, we run $6,000$ iterations of the MCMC algorithm and discard the first $1,000$ iterations as burn-in period. 
Table \ref{beta5dim} lists the posterior means of $p$ and the five-dimensional $\bm{\beta}$ of scenario 1 for the different values of $n$ and the true $p$, averaged over the $100$ simulated datasets. 
The figures in brackets show the standard deviations over the $100$ simulations. We notice that the posterior means of $p$ are close to their true values when $p$ is less than 1, corresponding to the ``heavy tail'' situation, but the estimation accuracy drops when $p$ is greater than 1. This similar performance is also confirmed for the posterior means of the vector of unknown coefficients $\bm{\beta} = (\beta_0,\beta_1,\beta_2,\beta_3,\beta_4)$.
Generally, increasing the sample size from $100$ to $250$ leads to better estimations of the model parameters.  
Table \ref{beta10dim} shows the posterior means and, in brackets, the standard deviations over the $100$ different simulated datasets, for the ten-dimensional scenario. These results are in line with the five-dimensional case, thus leading us to conclude that increasing the number of covariates is not an issue in terms of estimation accuracy. Moreover, most standard deviations in Tables \ref{beta5dim} and \ref{beta10dim} decrease for the different scenarios and $p$ values when we move from $n=100$ to $n=250$.

\noindent We further compare the performance of our P\'olya-Gamma sampler to the one of $BC_{el}$. Table \ref{EL:beta5dim} and Table \ref{EL:beta10dim} shows the posterior means of $p$ and $\bm{\beta}$, and the standard deviations over $100$ different simulated datasets. As in the scenario with $p$ known, we choose a much less vaguer normal prior distribution $\mathcal{N}(\bm{\nu}=0,B=9\cdot I_k)$ for $\bm{\beta}$. Results are based on 20,000 posterior samples obtained with $BC_{el}$. It can be easily seen that when $p<1$, the P\'olya-Gamma sampler outperforms the $BC_{el}$ substantially, in terms of both the posterior means and standard deviations across different datasets. The advantage of P\'olya-Gamma sampler against $BC_{el}$ reduces when $p>1$, however, the standard deviations still suggest that P\'olya-Gamma sampler is more consistent.
 
\begin{figure}[h!]
	\centering
	\begin{tabular}{ccc}
		{\includegraphics[width=4.75cm]{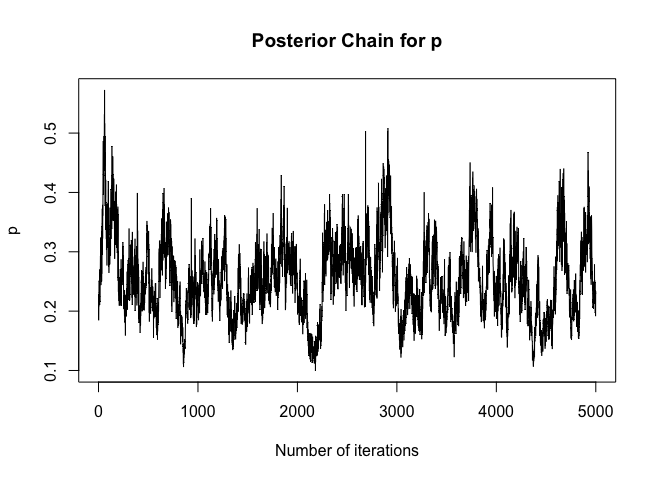}} &
		{\includegraphics[width=4.75cm]{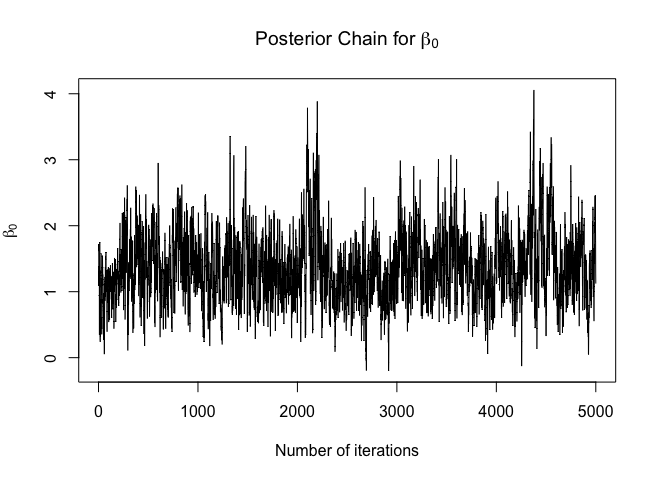}} &
		{\includegraphics[width=4.75cm]{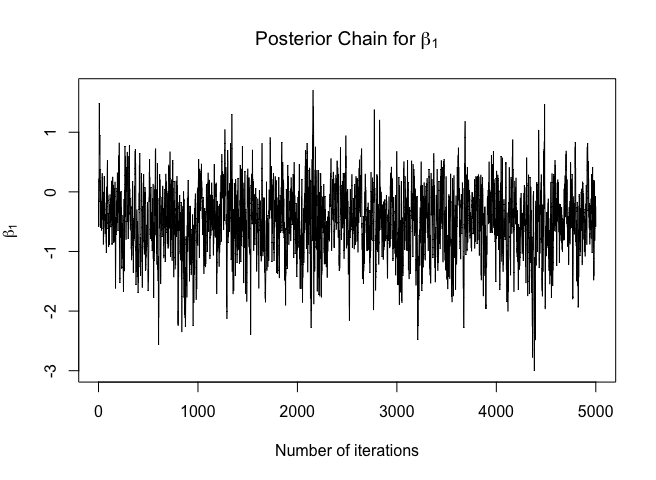}} \\
		{\includegraphics[width=4.75cm]{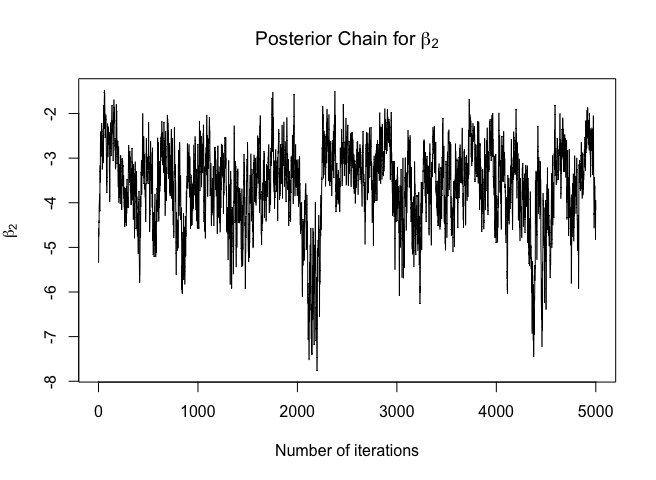}} &
		{\includegraphics[width=4.75cm]{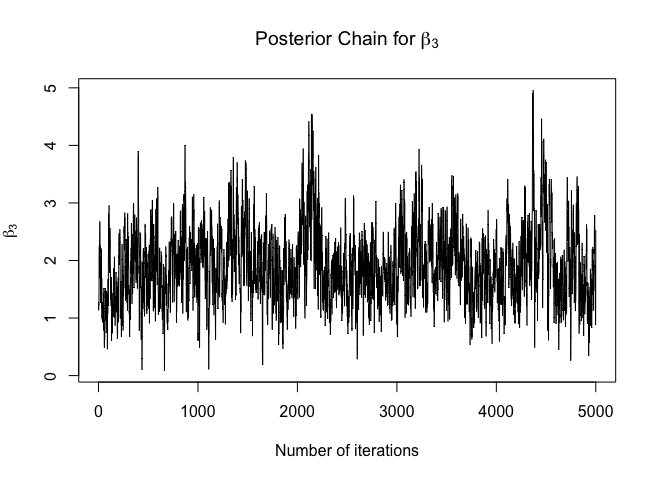}} &
		{\includegraphics[width=4.75cm]{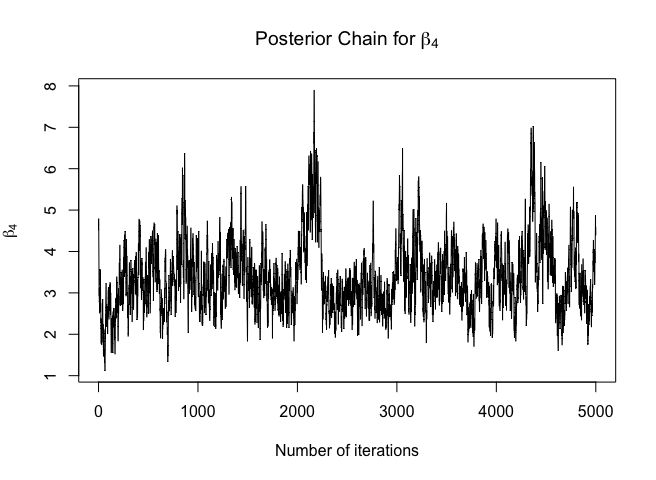}} \\
	\end{tabular}
	\caption{Case with $p$ unknown: sample chains of the posterior distribution of the parameters for the simulated data with sample size $n=250$ and with true parameter values $p=0.3$ and $\bm{\beta} = (\beta_0,\beta_1,\beta_2,\beta_3,\beta_4) = (1,-1,-3,1,3)$.}
	\label{Fig_Chain_Simu}
\end{figure}

 \noindent As done previously with $p$ known, here, with $p$ unknown, we display in Figure \ref{Fig_Chain_Simu} the posterior chains for a specific dataset with sample size $n=250$, where $p=0.3$ and $\bm{\beta} = (\beta_0,\beta_1,\beta_2,\beta_3,\beta_4) = (1,-1,-3,1,3)$.
Despite exhibiting slightly slower mixing behaviours compared to Figure \ref{Fig_Chain_Simu_fix},
the chains, after a burn-in of $1,000$ iterations, converge relatively quickly to the true values of all parameters. 
\noindent We also computed Geweke's convergence tests for the values of $p$ and five-dimensional $\bm{\beta}$, with $n=100$ and $n=250$ and $p=0.3${\footnote{Results for the  ten-dimensional case are omitted for lack of space but are available on request.}}. Table \ref{Tab_Gew} reports the results of Geweke's convergence test \citep{Geweke1992}, that indicate no issues since the absolute values of the test statistics are all lower than $1.96$.

\begin{table}[htbp]
	\centering
	\begin{tabular}{lcccccc}
		\hline
		& $p$ &  $\beta_0$ & $\beta_1$ & $\beta_2$ &  $\beta_3$ &  $\beta_4$  \\
		\hline
		\emph{true values} & 0.3 & 1 & -1 & -3 & 1 & 3 \\
		\hline
		$n = 100$   & 0.271   & 1.400 &	-1.311& 	-3.882& 	1.330& 	3.946   \\
		& (0.096) & (0.815) & (0.667) & (0.629) & (0.828) & (0.689) \\
		$n = 250$  & 0.230  & 1.439 &	-1.480& 	-4.341& 	1.442& 	4.489  \\
		& (0.056) & (0.553) & (0.583) & (0.565) & (0.599) & (0.635) \\
		\hline
		\emph{true values} & 0.7 & 1 & -1 & -3 & 1 & 3 \\
		\hline
		$n = 100$   & 0.798 & 0.995 & -0.976 & -2.991 & 1.028 & 2.884 \\
		& (0.174) & (0.320) & (0.394) & (0.383) & (0.393) & (0.384) \\
		$n = 250$   & 0.821 & 0.927 & -0.959 & -3.026 &  0.966 & 3.117\\
		& (0.214) & (0.271) & (0.283) & (0.345) & (0.287) & (0.345) \\
		\hline
		\emph{true values} & 1.5 & 1 & -1 & -3 & 1 & 3 \\
		\hline
		$n = 100$   & 1.129  & 1.388& 	-1.390 &	-4.335 &	1.427 &	4.351  \\
		& (0.351) & (0.382) & (0.405) & (0.526) & (0.528) & (0.490) \\
		$n = 250$   & 1.164  & 1.465& 	-1.436& 	-4.381& 	1.490& 	4.444 \\
		& (0.265) & (0.318) & (0.307) & (0.579) & (0.382) & (0.557) \\
		\hline
		\emph{true values} & 3 & 1 & -1 & -3 & 1 & 3 \\ 
		\hline
		$n = 100$   &  1.593  & 1.506 &	-1.576& 	-4.697& 	1.607& 	4.736 \\
		& (0.387) & (0.415) & (0.424) & (0.563) & (0.528) & (0.608) \\
		$n = 250$   & 1.691  & 1.570 &	-1.607& 	-4.778& 	1.560& 	4.740 \\
		& (0.412) & (0.267) & (0.324) & (0.587) & (0.289) & (0.586) \\
		\hline
	\end{tabular}
	\caption{PG sampler. Case with $p$ unknown: posterior means over the $100$ different simulated datasets for $n$ equal to $100$ and $250$ compared with the true values of $p = 0.3$ (first panel), $p=0.7$ (second panel), $p=1.5$ (third panel) and $p=3$ (fourth panel) and $(\beta_0,\beta_1,\beta_2,\beta_3,\beta_4) = (1,-1,-3,1,3)$. The values in brackets are the standard deviations over the $100$ different simulations.}
	\label{beta5dim}
\end{table}

\begin{table}[htbp]
	\centering
	\begin{tabular}{lcccccc}
		\hline
		& $p$ &  $\beta_0$ & $\beta_1$ & $\beta_2$ &  $\beta_3$ &  $\beta_4$  \\
		\hline
		\emph{true values} & 0.3 & 1 & -1 & -3 & 1 & 3 \\
		\hline
		$n = 100$   & 0.559    & 0.258& 	-0.264& 	-0.868& 	0.345& 	0.930    \\
		& (0.520) & (0.440) & (0.394) & (0.577) & (0.480) & (0.631) \\
		$n = 250$  &0.149 & 0.166 &	-0.165 &	-0.427& 	0.139 &	0.398   \\
		& (0.269) & (0.507) & (0.445) & (0.603) & (0.484) & (0.548) \\
		\hline
		\emph{true values} & 0.7 & 1 & -1 & -3 & 1 & 3 \\
		\hline
		$n = 100$   & 0.807 & 0.435 & -0.576 & -1.991 & 0.498 & 1.579 \\
		& (0.698) & (0.507) & (0.438) & (0.570) & (0.491) & (0.581) \\
		$n = 250$   & 0.754 & 0.483 & -0.591 & -2.076 &  0.514 & 1.618\\
		& (0.214) & (0.471) & (0.414) & (0.610) & (0.437) & (0.519) \\
		\hline
		\emph{true values} & 1.5 & 1 & -1 & -3 & 1 & 3 \\
		\hline
		$n = 100$   & 2.464  & 0.563& 	-0.617 &	-1.626& 	0.595 &	1.716   \\
		& (1.378) & (0.558) & (0.494) & (0.489) & (0.441) & (0.556) \\
		$n = 250$   & 1.877  & 0.271& 	-0.483& 	-1.619 &	0.377 &	1.518  \\
		& (1.687) & (1.013) & (0.629) & (0.774) & (0.583) & (0.787) \\
		\hline
		\emph{true values} & 3 & 1 & -1 & -3 & 1 & 3 \\ 
		\hline
		$n = 100$   &  3.056  & 0.632 &	-0.391& 	-1.385 &	0.508 &	1.433  \\
		& (1.204) & (0.630) & (0.601) & (0.857) & (0.551) & (0.945) \\
		$n = 250$   & 2.641   & 0.478 &	-0.551 &	-1.570& 	0.598 &	1.634  \\
		& (1.759) & (0.750) & (0.705) & (0.728) & (0.594) & (0.723) \\
		\hline
	\end{tabular}
	\caption{$BC_{el}$ sampler. Case with $p$ unknown: posterior means over the $100$ different simulated datasets for $n$ equal to $100$ and $250$ compared with the true values of $p = 0.3$ (first panel), $p=0.7$ (second panel), $p=1.5$ (third panel) and $p=3$ (fourth panel) and $(\beta_0,\beta_1,\beta_2,\beta_3,\beta_4) = (1,-1,-3,1,3)$. The values in brackets are the standard deviations over the $100$ different simulations.}
	\label{EL:beta5dim}
\end{table}

\begin{sidewaystable}[htbp]
		\centering
		\begin{tabular}{lccccccccccc}
			\hline
			& $p$ &  $\beta_0$ & $\beta_1$ & $\beta_2$ &  $\beta_3$ &  $\beta_4$ &  $\beta_5$ &  $\beta_6$ &  $\beta_7$ &  $\beta_8$ &  $\beta_9$   \\
			\hline
			\emph{true values} & 0.3 & 2.3 & 1 & -2 & 1.5 & -2.7 & 0.2 & -1.4 & 3 & -0.6 & -1.2\\
			\hline
			$n = 100$   & 0.197  & 3.440 &	1.395& 	-3.179& 	2.271& 	-3.667& 	0.097& 	-2.093& 	4.356& 	-1.044 &	-1.820  \\
			& (0.063) & (0.735)& (1.221)& (1.128)& (0.903)& (0.868)& (1.126)& (1.025)& (1.031)& (0.948)& (0.959) \\
			$n = 250$  & 0.186  & 3.685 &	1.656& 	-3.341& 	2.563& 	-4.497& 	0.447& 	-2.309& 	4.982& 	-1.070& 	-1.875 \\
			& (0.040) & (0.696)& (0.648)& (0.657)& (0.706)& (0.678)& (0.757)& (0.667)& (0.722)& (0.839)& (0.646) \\
			\hline
			\emph{true values} & 0.7 & 2.3 & 1 & -2 & 1.5 & -2.7 & 0.2 & -1.4 & 3 & -0.6 & -1.2\\
			\hline
			$n = 100$ & 0.658 & 2.517&  1.132& -2.213&  1.739& -3.016&  0.202& -1.608&  3.264& -0.713& -1.112  \\
			& (0.197) & (0.644)& (0.673)& (0.550)& (0.543)& (0.465)& (0.575)& (0.617)& (0.654)& (0.642)& (0.650) \\
			$n = 250$  & 0.657 &  2.653&  1.186& -2.286&  1.883& -3.172&  0.248& -1.593&  3.410& -0.693& -1.408 \\
			& (0.138) & (0.306)& (0.443)& (0.448)& (0.435)& (0.396)& (0.375)& (0.289)& (0.494)& (0.285)& (0.303) \\
			\hline
			\emph{true values} & 1.5 & 2.3 & 1 & -2 & 1.5 & -2.7 & 0.2 & -1.4 & 3 & -0.6 & -1.2\\
			\hline
			$n = 100$  & 0.956  & 3.678 &	1.657& 	-3.351& 	2.397& 	-4.363& 	0.316& 	-2.286& 	4.876& 	-1.014& 	-1.974  \\
			& (0.329) & (0.561)& (0.611)& (0.548)& (0.665)& (0.672)& (0.650)& (0.697)& (0.680)& (0.645)& (0.578) \\
			$n = 250$  & 0.986  & 3.818 &	1.611& 	-3.306& 	2.460& 	-4.423& 	0.339& 	-2.362& 	4.913& 	-0.921& 	-1.940 \\
			& (0.263) & (0.430)& (0.372)& (0.407)& (0.408)& (0.486)& (0.358)& (0.385)& (0.428)& (0.361)& (0.405) \\
			\hline
			\emph{true values} & 3 & 2.3 & 1 & -2 & 1.5 & -2.7 & 0.2 & -1.4 & 3 & -0.6 & -1.2\\ 
			\hline
			$n = 100$  & 1.471  & 3.804 &	1.731& 	-3.314& 	2.389& 	-4.510& 	0.308& 	-2.497& 	5.029& 	-1.156& 	-2.014  \\
			& (0.378) & (0.533)& (0.590)& (0.595)& (0.646)& (0.603)& (0.568)& (0.595)& (0.674)& (0.517)& (0.482) \\
			$n = 250$  & 1.488  & 4.008 &	1.656& 	-3.449& 	2.595& 	-4.600& 	0.387& 	-2.426& 	5.146& 	-1.042 &	-2.054 \\
			& (0.430) & (0.540)& (0.358)& (0.480)& (0.441)& (0.603)& (0.338)& (0.434)& (0.655)& (0.326)& (0.373) \\
			\hline
		\end{tabular}
		\caption{PG sampler. Case with $p$ unknown: posterior means over the $100$ different simulated datasets for $n$ equal to $100$ and $250$ compared with the true values of $p = 0.3$ (first panel), $p=0.7$ (second panel), $p=1.5$ (third panel) and $p=3$ (fourth panel) and $(\beta_0,\beta_1,\beta_2,\beta_3,\beta_4,\beta_5,\beta_6,\beta_7,\beta_8,\beta_9) = (2.3,1,-2,1.5,-2.7,0.2,-1.4,3,-0.6,-1.2)$. The values in brackets are the standard deviations over the $100$ different simulations.}
		\label{beta10dim}
\end{sidewaystable}

\begin{sidewaystable}[htbp]
	\centering
	\begin{tabular}{lccccccccccc}
		\hline
		& $p$ &  $\beta_0$ & $\beta_1$ & $\beta_2$ &  $\beta_3$ &  $\beta_4$ &  $\beta_5$ &  $\beta_6$ &  $\beta_7$ &  $\beta_8$ &  $\beta_9$   \\
		\hline
		\emph{true values} & 0.3 & 2.3 & 1 & -2 & 1.5 & -2.7 & 0.2 & -1.4 & 3 & -0.6 & -1.2\\
		\hline
		$n = 100$   & 0.149   & 0.554 &	0.300 &	-0.697& 	0.321 &	-0.690 &	0.207& 	-0.412 &	0.844 &	-0.232& 	-0.348  \\
		& (0.164) & (0.590)& (0.608)& (0.831)& (0.568)& (0.613)& (0.582)& (0.657)& (0.629)& (0.671)& (0.609) \\
		$n = 250$  & 0.056   & 0.409 &	0.128& 	-0.334& 	0.242 &	-0.478 &	0.040& 	-0.184& 	0.493& 	-0.208& 	-0.178  \\
		& (0.085) & (0.535)& (0.402)& (0.470)& (0.497)& (0.560)& (0.502)& (0.583)& (0.566)& (0.663)& (0.485) \\
		\hline
		\emph{true values} & 0.7 & 2.3 & 1 & -2 & 1.5 & -2.7 & 0.2 & -1.4 & 3 & -0.6 & -1.2\\
		\hline
		$n = 100$ & 0.885 & 0.847&  0.385& -0.573&  0.418& -0.650&  0.120& -0.401&  0.710& -0.258& -0.307  \\
		& (0.286) & (0.604)& (0.649)& (0.541)& (0.589)& (0.494)& (0.580)& (0.682)& (0.697)& (0.657)& (0.662) \\
		$n = 250$  & 0.641 &  0.896&  0.401& -0.580&  0.515& -0.671&  0.158& -0.534&  0.790& -0.277& -0.401 \\
		& (0.288) & (0.708)& (0.749)& (0.701)& (0.794)& (0.755)& (0.801)& (0.583)& (0.651)& (0.497)& (0.724) \\
		\hline
		\emph{true values} & 1.5 & 2.3 & 1 & -2 & 1.5 & -2.7 & 0.2 & -1.4 & 3 & -0.6 & -1.2\\
		\hline
		$n = 100$  & 2.228  & 0.955 &	0.397 &	-0.454& 	0.434 &	-0.631 &	-0.025& 	-0.382& 	0.617& 	-0.257& 	-0.147   \\
		& (2.090) & (0.811)& (0.773)& (1.079)& (0.983)& (1.121)& (0.980)& (0.926)& (1.363)& (0.743)& (0.986) \\
		$n = 250$  & 1.118   & 1.191 &	0.423 &	-0.738& 	0.595& 	-0.773 &	0.085 &	-0.525& 	1.284 &	-0.303 &	-0.368  \\
		& (1.543) & (0.758)& (0.885)& (0.948)& (0.939)& (1.030)& (0.970)& (0.976)& (1.111)& (0.958)& (0.925) \\
		\hline
		\emph{true values} & 3 & 2.3 & 1 & -2 & 1.5 & -2.7 & 0.2 & -1.4 & 3 & -0.6 & -1.2\\ 
		\hline
		$n = 100$  & 3.337   & 0.567 &	0.055 &	-0.004& 	0.149 &	0.002 &	-0.028 &	-0.198 &	0.005 &	-0.018 &	-0.009   \\
		& (1.909) & (1.020)& (1.182)& (1.252)& (1.053)& (1.269)& (0.983)& (1.246)& (1.276)& (0.976)& (1.210) \\
		$n = 250$  & 2.041   & 0.995 &	0.148 &	-0.376 &	0.258 &	-0.379 &	0.186 &	-0.249 &	0.617& 	-0.110& 	-0.216  \\
		& (1.912) & (0.827)& (1.118)& (1.227)& (1.087)& (1.294)& (0.891)& (1.070)& (1.231)& (0.932)& (1.144) \\
		\hline
	\end{tabular}
	\caption{$BC_{el}$ sampler. Case with $p$ unknown: posterior means over the $100$ different simulated datasets for $n$ equal to $100$ and $250$ compared with the true values of $p = 0.3$ (first panel), $p=0.7$ (second panel), $p=1.5$ (third panel) and $p=3$ (fourth panel) and $(\beta_0,\beta_1,\beta_2,\beta_3,\beta_4,\beta_5,\beta_6,\beta_7,\beta_8,\beta_9) = (2.3,1,-2,1.5,-2.7,0.2,-1.4,3,-0.6,-1.2)$. The values in brackets are the standard deviations over the $100$ different simulations.}
	\label{EL:beta10dim}
\end{sidewaystable}

\begin{table}[h!]
	\centering
	\begin{tabular}{llcc|llc}
		\hline
		& Variables & Test Statistic & &  & Variables & Test Statistic\\
		\hline
		$n = 100$ & $p = 0.3$ & -0.507 & & $n = 250$ & $p = 0.3$ & 0.914 \\
		& $\beta_0 = 1$ & 1.721 & & & $\beta_0 = 1$ & -1.394 \\
		& $\beta_1 = -1$ & -1.308 & & & $\beta_1 = -1$ & 1.320\\
		& $\beta_2 = -3$ & -1.293 & & & $\beta_2 = -3$ &  1.321 \\
		& $\beta_3 = 1$ & 1.413 & & & $\beta_3 = 1$ &   -1.545 \\
		& $\beta_4 = 3$ & 0.539 & & & $\beta_4 = 3$ &  -1.259 \\
		\hline
	\end{tabular}
	\caption{Case with $p$ unknown: Geweke's test statistics for the posterior chain of the tail parameter $p$ and of the unknown coefficients, $\bm{\beta}= (\beta_0,\beta_1,\beta_2,\beta_3,\beta_4)$ for a simulated dataset of sample size $n=100$ (left panel) and $n=250$ (right panel).}
	\label{Tab_Gew}
\end{table}
\section{Real Data Applications} \label{Real}
In this Section, we present two different real data applications to illustrate the accurate and efficient performance of the P\'olya-Gamma sampler for estimating the generalized logistic regression.
The applications analyse novel datasets of different dimensions, which are related to current and topical problems in the social sciences.
In the first application, we consider data selected from the European Social Survey (ESS), that were previously analysed in \cite{OurArticle2019} 
to identify the determinants of public opinions towards immigration. 
The second application analyses the motives behind the recidivism of offenders released from prison between 2016 and 2018 in the State of Iowa, USA. 

\subsection{Immigration in European Countries}

The first dataset focuses on people's attitudes towards immigration. 
The data was selected from the ESS and was collected following hour-long face-to-face interviews.
We focused on questions regarding immigration from the ESS8 edition 1.0 published in October 2017 and collected between August 2016 and March 2017 \citep{ESS16, ESS18} for Great Britain (GB), Germany (DE) and France (FR), respectively.
For these three countries, the total number of observations is $5,354$, of which $1,419$ are for GB; $2,284$ for DE and $1,651$ for FR. 

The dependent variable is \texttt{immig}, that indicates whether the respondent would allow immigrants from poorer countries outside Europe. More precisely, if $\texttt{immig} = 1$ the respondent is against immigration, if $\texttt{immig} = 0$ the respondent is in favor of immigration. 
Following \cite{OurArticle2019}, we considered as covariates the variables: \texttt{pplfair}, \texttt{trstep}, \texttt{trstun}, \texttt{happy}, \texttt{agea}, \texttt{edulvlb} and \texttt{hinctnta}. The first three variables include answers to question ranging from $0$ (most negative) to $10$ (most positive): do you think that most people try to take advantage of you (\texttt{pplfair})? Do you trust the European parliament (\texttt{trstep})? Do you trust the United Nations (\texttt{trstun})? The \texttt{happy} variable takes values from $0$ (unhappy) to $10$ (extremely happy). The last three variables are subject-specific and include the age, ranging from $15$ to $100$ (\texttt{agea}); the highest level of education, from primary education to doctoral degree (\texttt{edulvlb}) and the household's total net income (\texttt{hinctnta}).

For this dataset, we run $5,000$ iterations of the P\'olya-Gamma sampler algorithm and we discarded the first $1,000$ iterations as burn-in. We used the MLE estimates of $\bm{\beta}$ and $p$ as starting points of the algorithm and we employed a normal prior with mean 0, standard deviation 5 times greater than the one estimated with a standard logistic regression.

In order to assess the performance of the P\'olya-Gamma sampler, we report the results obtained using the empirical likelihood approach on the same dataset, adopting a similar setting.
Table \ref{Tab_Immig} lists the results of the P\'olya-Gamma sampler (left-hand side) and the empirical likelihood approach (right-hand side).
For each European country (GB, DE and FR) and for each parameter, Table \ref{Tab_Immig} shows the posterior means, the 2.5\% and 97.5\% quantiles of the posterior distributions of the P\'olya-Gamma and the empirical likelihood methods.
The effects of the parameters are similar across countries and suggest that the higher the level of education, the higher the probability for people to be favourable towards immigration. 
Also, trustful people and individuals with a strong confidence in the European parliament tend to be in favour of immigration.  
The posterior mean estimates obtained with the P\'olya-Gamma method are in line with those obtained with the empirical likelihood.
For some of the covariates, such as \texttt{pplfair} and \texttt{trstep}, the empirical likelihood approach assigns
posterior support to the value zero, while the corresponding P\'olya-Gamma credible intervals (CIs) do not include zero. 
In addition, both methods agree on the estimation of the $p$ parameter for all countries, suggesting heavy tails for the generalised logistic regression.
Therefore, the P\'olya-Gamma sampler outperforms the empirical likelihood method for the estimation of the generalised logistic regression, since it provides more precise parameter estimates as for the $\bm{\beta}$.

\begin{table}[htbp]
	\centering
	\scriptsize
	\begin{tabular}{|c|l|ccc|ccc|}
		\hline
		\multicolumn{1}{|r}{} &       & \multicolumn{3}{c|}{P\'olya-Gamma Sampler} & \multicolumn{3}{c|}{Empirical Likelihood}  \\
		\hline
		\multicolumn{1}{|l|}{Country  } & Parameters & \multicolumn{1}{l}{ post. mean } & \multicolumn{1}{l}{ 2.50\% } & \multicolumn{1}{c|}{  97.50\% } & \multicolumn{1}{l}{ post. mean } & \multicolumn{1}{l}{ 2.50\% } & \multicolumn{1}{c|}{  97.50\% }  \\
		\hline
		\multirow{9}[2]{*}{ GB  } &  const  & -0.8 &	-2.441&	0.541 & -0.4202 & -2.4377 & 1.4487   \\
		&   \texttt{pplfair} & -0.223&	-0.552&	-0.068 & -0.1822 & -0.3176 & 0.0113  \\
		&   \texttt{trstep} & -0.362&	-0.888&	-0.123 & -0.1731 & -0.3796 & 0.0447  \\
		&   \texttt{trstun} & -0.075&	-0.302&	0.046 &  -0.0372 & -0.2354 & 0.1409 \\
		&   \texttt{happy} & 0.047&	-0.099&	0.241 &  -0.0133 & -0.2892 & 0.1209 \\
		&   \texttt{agea} & 0.015&	-0.001&	0.04 &  0.0094 & -0.0139 & 0.0358  \\
		&   \texttt{edulvlb} & -0.004&	-0.009&	-0.001 & -0.0021 & -0.0039 & -0.0006  \\
		&       \texttt{hinctnta} & 0.017&	-0.105&	0.148 &-0.044 & -0.1856 & 0.1228  \\
		&  $p$    & 0.815&	0.2&	1.944 &  0.8218 & 0.4284 & 1.5696 \\
		\hline
		\multirow{9}[2]{*}{ DE  } &  const  &-1.152&	-2.648&	0.29 &  -1.1452 & -2.4617 & 0.6941 \\
		&   \texttt{pplfair} & -0.301&	-0.616&	-0.114 & -0.1612 & -0.3291 & -0.0093  \\
		&   \texttt{trstep} & -0.429&	-1&	-0.148 &-0.1524 & -0.2839 & -0.0107  \\
		&   \texttt{trstun} & -0.091&	-0.33&	0.085& -0.0112 & -0.2115 & 0.1847  \\
		&   \texttt{happy} & -0.169&	-0.46&	0.003 &-0.0873 & -0.2438 & 0.0659  \\
		&   \texttt{agea} &  0.036&	0.013&	0.077& 0.0124 & -0.0061 & 0.0375  \\
		&   \texttt{edulvlb} & 0&	-0.004&	0.002 & -0.0007 & -0.0026 & 0.0016 \\
		&       \texttt{hinctnta} & -0.2&	-0.533&	-0.042 & -0.1171 & -0.2352 & 0.0414 \\
		&  $p$    &0.608&	0.155&	1.814 & 0.7338 & 0.2955 & 1.7036 \\
		\hline
		\multirow{9}[2]{*}{ FR } &  const  &1.054&	-0.35&	3.361 & 0.8442 & -0.3443 & 1.4314 \\
		&   \texttt{pplfair} & -0.111&	-0.347&	0.01 &  -0.0921 & -0.1898 & 0.0193  \\
		&   \texttt{trstep} & -0.364&	-0.872&	-0.107 & -0.1912 & -0.3592 & -0.0114  \\
		&   \texttt{trstun} & -0.122&	-0.353&	0.004& -0.0514 & -0.2047 & 0.1417\\
		&   \texttt{happy} & 0.029&	-0.129&	0.227 &  0.0191 & -0.0609 & 0.1545 \\
		&   \texttt{agea} & 0.001&	-0.013&	0.021& -0.0043 & -0.0155 & 0.0091\\
		&   \texttt{edulvlb} & -0.005&	-0.012&	-0.002 & -0.0038 & -0.0056 & -0.0027 \\
		&       \texttt{hinctnta} & -0.092&	-0.275&	0.016 & -0.1061 & -0.239 & 0.0286 \\
		&  $p$    & 0.827&	0.159&	2.055 &0.7845 & 0.3991 & 1.2809  \\
		\hline
	\end{tabular}%
	\caption{EU immigration data results for Great Britain (GB), Germany (DE) and France (FR), obtained with the P\'olya-Gamma (left-hand side) and the empirical  likelihood (right-hand side) methods. For each parameter, columns 3 and 6 show the posterior means, columns 4 and 7 show the 2.5\% quantiles of the posterior distribution, columns 5 and 8 show the 97.5\% quantiles of the posterior distribution.}  \label{Tab_Immig}
\end{table}

\subsection{Recidivism of Offenders Released from Prison in Iowa}

The second dataset contains information about the recidivism of offenders released from prison between 2016 and 2018 in Iowa\footnote{See the following website for details: \href{https://data.iowa.gov/Correctional-System/3-Year-Recidivism-for-Offenders-Released-from-Pris/mw8r-vqy4}{https://data.iowa.gov/Correctional-System/3-Year-Recidivism-for-Offenders-Released-from-Pris/mw8r-vqy4}.}. Recidivism happens when a prisoner is released from jail and he or she relapses back into criminal behavior and returns to jail. 
We applied the generalised logistic regression to determine whether a prisoner is likely to recidivate based on his or her characteristics. 




After removing missing values, the data consists of $13,644$ records corresponding to offenders detained in prison and released from 2016 to 2018. 
The dependent variable is $\texttt{recidivism}$, which is equal to $1$ if the prisoner is arrested again within a $3$-year tracking period after being released, and it is equal to $0$ if he or she has not returned to prison within the tracking period.


We model the probability of being recidivist using a subsample of all the explanatory variables present in the original database: 
the sex of the offender (\texttt{sex}, which is equal to $1$ for men and $0$ for women); the age when released from prison (\texttt{age}); whether an offender committed a felony or misdemeanor (\texttt{felony}, which is equal to $1$ for felony and $0$ for misdemeanor); whether an offender was charged with a drug crime (\texttt{drug}); whether an offender was charged with a public order crime  (\texttt{puborder}); whether an offender was charged with a violent crime  (\texttt{violent}); and whether an offender was released on discharge or parole (\texttt{discharge}, which is equal to $1$ for discharge and $0$ for parole). 

In this second example, we run the MCMC algorithm using $5,000$ iterations and discarding the first $1,000$ iterations. As in the immigration example, we adopted vague priors and we chose the initial values of $\bm{\beta}$ and $p$ based on the corresponding MLE estimators.
%

\begin{table}[htbp]
	\centering
	\scriptsize
	\begin{tabular}{|l|ccc|ccc|}
		\hline
		&   \multicolumn{3}{c|}{P\'olya-Gamma Sampler} & \multicolumn{3}{c|}{Empirical Likelihood}  \\
		\hline
		Parameters & \multicolumn{1}{l}{ post. mean } & \multicolumn{1}{l}{ 2.50\% } & \multicolumn{1}{l|}{  97.50\% }  & \multicolumn{1}{l}{ post. mean } & \multicolumn{1}{l}{ 2.50\% } & \multicolumn{1}{l|}{  97.50\% }\\
		\hline
		const & -0.1443&	-0.3811&	0.0232  & -0.4278 & 	-0.9341 & 0.1719 \\
		\texttt{sex}& 0.311&	0.1468&	0.5873  & 0.1371 & -0.2583 & 0.4574\\
		\texttt{age}& -0.0004&	-0.0007&	-0.0002  & -0.0003 & -0.0005 & -0.0002 \\
		\texttt{felony}& -0.0941&	-0.2521&	0.0184 & -0.0974 & -0.4457 & 0.0999  \\
		\texttt{drug}& -0.0308&	-0.1496&	0.0774 & -0.0996  & -0.4009 & 0.1912\\
		\texttt{puborder}& -0.2769&	-0.5663&	-0.116  & -0.0694 & -0.3384 & 0.2261 \\
		\texttt{violent}& -0.5759&	-1.0713&	-0.3271  & -0.4743 & -0.9779 & -0.2638 \\
		\texttt{discharge}& -0.5736&	-1.068&	-0.3225  & -0.4654 & -0.6726 & -0.1936 \\
		$p$ & 0.9343&	0.3088&	1.8439 & 0.5679 & 0.3409 & 1.0905 \\
		\hline  
	\end{tabular}
	\caption{Criminal recidivism data results obtained with the P\'olya-Gamma (left-hand side) and the empirical  likelihood (right-hand side) methods. For each parameter, columns 2 and 5 show the posterior means, columns 3 and 6 show the 2.5\% quantiles of the posterior distribution, columns 4 and 7 show the 97.5\% quantiles of the posterior distribution.} \label{Tab_Recid}
\end{table}

As we did with the immigration data, we compared the performances of the P\'olya-Gamma sampler and the empirical likelihood approaches, using the same settings for both algorithms.
The results for the parameters $\bm{\beta}$ and $p$ are listed in Table \ref{Tab_Recid}, that displays the P\'olya-Gamma results on the left-hand side and the empirical likelihood results on the right-hand side. Table \ref{Tab_Recid} illustrates the posterior means, the 2.5\% and 97.5\% quantiles of the posterior distributions of the P\'olya-Gamma and the empirical likelihood methods.

The results suggest that men are more likely to recidivate than women, younger prisoners are more likely to become repeat offenders than older prisoners, there is a lower probability that the criminal is a recidivist if he or she was charged with a public order or violent crime and offenders released on parole are more likely to return to jail than those released on discharge.

The sign of the posterior means for both approaches is the same, suggesting a similar direction for the effects obtained using the P\'olya-Gamma and the empirical likelihood method. However, the CIs of the first approach are narrower than those of second approach.
Moreover, the P\'olya-Gamma method, unlike the empirical likelihood, does not give posterior support to zero to the variables \texttt{sex} and \texttt{puborder}, suggesting that the inclusion of these variables in the model is appropriate.


In agreement with our findings with the immigration data, the results obtained with the recidivism dataset demonstrate that the P\'olya-Gamma sampler method yields a higher estimates precision and accuracy compared to the empirical likelihood method for the covariate parameters $\bm{\beta}$.

\section{Conclusions} \label{Conclusion}

This paper introduces a novel DA scheme for the generalized logistic regression model. 
This model is particularly flexible since it is able to accommodate both light and heavy tails in dichotomous response data. 
The proposed DA scheme makes use of the P\'olya-Gamma identity and it is strongly related to the slice sampler algorithm.  
The P\'olya-Gamma sampler allows us to implement a Bayesian method able to draw samples from the exact posterior distribution.
On the contrary, other methods, such as the empirical likelihood approach, are based on approximations of the posterior and may lead to low precision in the estimation of the parameters.
Our simulation study demonstrates the estimation accuracy of the newly proposed P\'olya-Gamma sampler with datasets of different dimensions. 
We compared the performances of the P\'olya-Gamma and the empirical likelihood methods for modelling new interesting datasets regarding the opinion on immigration in European countries and the probability of being recidivist for prisoners in Iowa, USA.
Our results demonstrate the superiority of the P\'olya-Gamma sampler over the empirical likelihood in terms of parameter precision.
The P\'olya-Gamma method allows a more accurate estimation of the tail parameter for the generalised logistic regression compared to approximate methods.

\bibliographystyle{apalike}

\bibliography{PolyaGamma}
\end{document}